# Determination of the electric field gradient and the magnetic field in Mössbauer spectrocsopy by half-cube method

Krzysztof R. Szymański

University of Białystok, Faculty of Physics, Ciołkowskiego 1L, 15-245 Białystok, Poland

Abstract

This paper presents a new method to determine all components of the electric field gradient tensor and orientation of the hyperfine magnetic field axis in the absorber Cartesian frame for Mössbauer spectroscopy for nuclear transitions between levels with spin 3/2 and 1/2. The method can be applied for single-crystal absorbers with well separated absorption lines in their spectra. Explicit formulas derived from velocity moments are presented. The new method allows full separation of the electric quadrupolar and magnetic dipolar hyperfine interactions by using unpolarized radiation.

## 1. Introduction

Determining the hyperfine parameters from the Mössbauer spectrum has been widely discussed in the literature, and it is known that all parameters cannot be determined from a texture-free spectra [1]. Only three independent invariants and the hyperfine magnetic field (*hmf*) can be determined. How to determine all parameters from the spectrum of a single crystal by using polarized radiation has been presented [2, 3]; however, using unpolarized radiation has not been fully discussed yet.

We show that all components of the electric field gradient (*efg*) and orientation of the *hmf* axis can be determined from measurements on a single crystal with a few different directions of the wave vector of the photon with respect to the absorber frame. The explicit form of the intensity tensor [3], introduced by Zimmermann [4, 5], and the concept of velocity moments [6] can be used to separate *efg* and *hmf*.

This paper is organized as follows. First, we briefly present some earlier results related to the intensity tensor, velocity spectra, and invariants used in this study. Next, we show how to separate *efg* and *hmf* in single site, single-crystal absorber and how to measure these tensors with respect to the absorber frame. We also present a special holder useful in laboratory practice.

## 2. Theoretical background

For mixed interactions, that is, magnetic dipole and electric quadrupole, the explicit form of the Mössbauer spectrum for nuclear spin in excited state $I_e = 3/2$ and ground state $I_g = 1/2$ can be obtained in the formalism of the intensity tensor [3]. The intensity tensor $I_{\alpha\beta}$ is constructed from the *efg* $V$ and the *hmf* pseudovector $B$. Index $\alpha$ corresponds to four excited states ($\alpha = 1,2,3,4$) while $\beta$ to the ground nuclear states ($\beta = \pm 1$). Tr$I_{\alpha\beta}$, $I^s_{\alpha\beta}$, and $G_{\alpha\beta}$ are the trace, symmetric, and antisymmetric parts of the intensity tensor, respectively. In the thin-absorber approximation, the area under an absorption line for circular polarization is equal to [3, 6, 7]:

$$A_{\alpha\beta\zeta} = f_s t \frac{\pi\Gamma}{2}\frac{1}{2}\left(\mathrm{Tr}\,\boldsymbol{I}_{\alpha\beta} - \boldsymbol{\gamma}\cdot\boldsymbol{I}^s_{\alpha\beta}\cdot\boldsymbol{\gamma} - 2\zeta\,\boldsymbol{G}_{\alpha\beta}\cdot\boldsymbol{\gamma}\right), \qquad (1)$$

where $t$ is the effective thickness of the absorber, $f_s$ is the recoilless fraction of radiation emitted by the source, $\Gamma$ is the natural width, and $\boldsymbol{\gamma}$ is a unit vector parallel to the direction of the photon. Index $\zeta = \pm 1$ describes two opposite circular polarizations of the photon. Absorption line positions $v_{\alpha\beta}$ (in velocity units) depend on the eigenvalues $\lambda_\alpha$ of spin 3/2 Hamiltonian:

$$v_{\alpha\beta} = \frac{B}{2}\left(\gamma_{3/2}\lambda_\alpha - \beta\gamma_{1/2}\right) + \delta, \qquad (2)$$

where $\delta$ is the isomer shift. Values $\lambda_\alpha$ are proportional to the excited state energy levels, $E_\alpha$ (in energy units) $\lambda_\alpha = 2E_\alpha/(\gamma_{3/2}B)$, where $\gamma_I = g_I\mu_N c/E$, $g_I$ is a nuclear g-factor, $\mu_N$ is the nuclear magneton, $E$ is the energy between the ground and the excited nuclear states, and $c$ is the speed of light.

For nucleus in a single site and with mixed interactions, shape of the spectrum is fully defined by the *efg* components, orientation of the *hmf*, and orientation of the *k* vector of the photon (see Fig. 1*a*). Shape of the spectrum means the line intensities $A_{\alpha\beta\zeta}$ and line positions $v_{\alpha\beta}$.

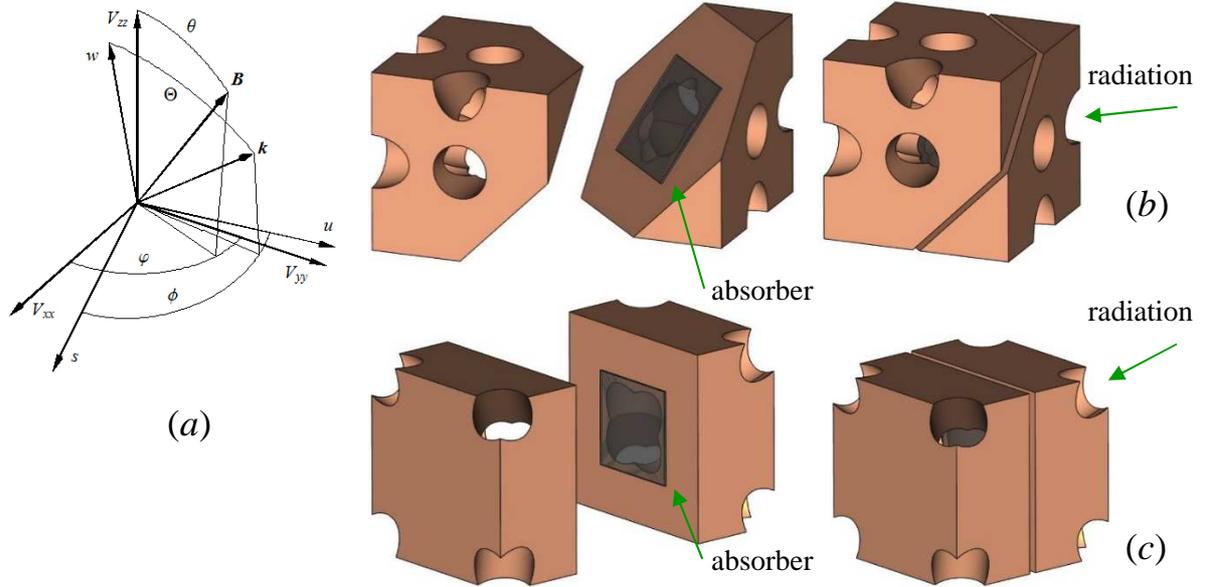

Fig. 1 (*a*) *PAS* of an electric field gradient ($V_{xx}$, $V_{yy}$, $V_{zz}$) and an absorber frame axes ($s$, $u$, $w$); wave vector of photon (*k*) and *hmf* (*B*) are also shown. Two kinds of sample holders are shown in (*b*) and (*c*). Each holder is a cube divided by two halves. A flat absorber is mounted between the halves at the center of the cube. Radiation passes through the holes on the edges, faces, or vertices.

The *n*th velocity moment of the spectrum, having dimension of velocity to the power of *n* is defined as:

$$W^n_\zeta = p\cdot\sum_{\alpha\beta} v^n_{\alpha\beta} A_{\alpha\beta\zeta}, \qquad p^{-1} = \sum_{\alpha\beta} A_{\alpha\beta\zeta}. \qquad (6)$$

The explicit form of the moments of the spectra were given in [6] and for the purposes of our analysis we need only:

$$W_\zeta^1 = \delta + \frac{1}{8} a\boldsymbol{\gamma} \cdot \boldsymbol{V} \cdot \boldsymbol{\gamma} - \zeta \frac{1}{4}(\gamma_{1/2} - 5\gamma_{3/2})B\,\boldsymbol{m} \cdot \boldsymbol{\gamma}, \qquad (7)$$

$$W_0^2 = \frac{1}{4}\left(\gamma_{1/2}^2 - 3\gamma_{1/2}\gamma_{3/2} + 4\gamma_{3/2}^2\right)B^2 - \frac{1}{4}\gamma_{3/2}(\gamma_{1/2} - 3\gamma_{3/2})B^2 (\boldsymbol{m} \cdot \boldsymbol{\gamma})^2$$
$$+ \frac{1}{4} a\delta\boldsymbol{\gamma} \cdot \boldsymbol{V} \cdot \boldsymbol{\gamma} + \delta^2 + \frac{1}{24} a\mathrm{Tr}\boldsymbol{V}^2, \qquad (8)$$

where $a = eQc/E$ is a proportionality constant between the *efg* components (in [V/m$^2$] SI units) and the velocity (in [mm/s]). $Q$ is the nuclear quadrupole moment, and $e$ is the elementary charge (positive value).

3. **Separation of the electric field gradient, hyperfine magnetic field, and the isomer shift**

The essence of equations (7, 8) is that the left-hand sides can be easily measured from the Mössbauer spectra. The right-hand sides represent interactions of nuclear spin and electromagnetic static fields and also some experimental conditions—orientation of a $\boldsymbol{k}$ vector with respect to the *efg* frame.

Equations (7,8) can be solved, and explicit form of scalars constructed from the *efg* $\boldsymbol{V}$, *hmf* $\boldsymbol{B}$ ($\boldsymbol{m} = \boldsymbol{B}/B$), and the $\boldsymbol{k}$ ($\boldsymbol{\gamma} = \boldsymbol{k}/k$) vector of the photon can be found:

$$\delta = W_{tf}^1, \qquad (9)$$

$$(\boldsymbol{m} \cdot \boldsymbol{\gamma})^2 = \frac{1}{3} - \frac{4}{\gamma_{3/2}(\gamma_{1/2} - 3\gamma_{3/2})B^2}\left(W_0^2 - W_{tf}^2 - 2W_{tf}^1\left(W_0^1 - W_{tf}^1\right)\right), \qquad (10)$$

$$\boldsymbol{m} \cdot \boldsymbol{\gamma} = \frac{2}{(\gamma_{1/2} - 5\gamma_{3/2})B}\left(W_{-1}^1 - W_1^1\right), \qquad (11)$$

$$\boldsymbol{\gamma} \cdot \boldsymbol{V} \cdot \boldsymbol{\gamma} = \frac{8}{a}\left(W_0^1 - W_{tf}^1\right), \qquad (12)$$

where $W_{tf}^n$ is the *n*th velocity moment for a texture-free absorber, while $W_0^n = (W_{-1}^n + W_{+1}^n)/2$ is the *n*th velocity moment measured with unpolarized radiation. The orientations of *efg* and *hmf* in the absorber reference frame can be fully determined from the explicit form of (9–12). We have to point out that using unpolarized radiation from (10), we can determine orientation of the *hmf* axis (sign cannot be determined) while orientation of the axis and sign of *hmf* can be determined from (11) using polarized radiation.

The values and orientation of the hyperfine interactions in the absorber frame can be determined explicitly from a relatively small number of measurements. By three measurements of $\boldsymbol{m}\boldsymbol{\gamma}$, with circularly polarized radiation and with $\boldsymbol{\gamma}$ parallel to *x*, *y*, and *z* axes defining Cartesian coordinates in the absorber frame, we determine three components of vector $\boldsymbol{m}$.

Isomer shift is just the first moment of the texture-free absorber.

Length of the magnetic field $B$ can be uniquely determined from the ground state splitting $I_g = 1/2$. In practice, texture-free spectra can be analyzed for this purpose using invariants. In this analysis as the number of unknowns is equal to the number of degrees of freedom in the spectrum, one can avoid the problem of ambiguity [8, 9].

Before going into a detailed analysis of angular dependences of $(\boldsymbol{m}\boldsymbol{\gamma})^2$ and $\boldsymbol{\gamma}\cdot\boldsymbol{V}\cdot\boldsymbol{\gamma}$, we analyze a general case of a symmetric, real tensor $\boldsymbol{A}$. By measuring $\boldsymbol{\gamma}\cdot\boldsymbol{A}\cdot\boldsymbol{\gamma}$ for different orientations of $\gamma$, we obtain the combinations of components of $A$ in the absorber frame. Three measurements with $\gamma$ parallel to $s$, $u$, and $w$ axes, allows one to get $A_{ss}$, $A_{uu}$, and $A_{ww}$, respectively. By three other measurements with $\gamma$ parallel to (1,1,0), (1,0,1), and (0,1,1) directions, one gets $A_{ss}/2+A_{uu}/2+A_{su}$, $A_{ss}/2+A_{ww}/2+A_{sw}$, and $A_{uu}/2+A_{ww}/2+A_{uw}$, respectively. With these six measurements, all the six independent components of $A$ tensor in the absorber frame can be obtained. Further analysis is a solution of an eigenproblem for a real, symmetric three-dimensional matrix. For the special case of $(m\gamma)^2$, the solution is much simpler.

Since $(m\gamma)^2 = \gamma \cdot A \cdot \gamma$, where $A = m \otimes m$ ($A_{ij} = m_i m_j$), by measurements of $A_{su}$, $A_{sw}$, and $A_{uw}$ and solving the set of equations for unknowns $m_s$, $m_u$, $m_w$:

$$A_{su} = m_s m_u, \quad A_{sw} = m_s m_w, \quad A_{uw} = m_u m_w, \tag{13}$$

we get two solutions, $m = \pm(m_s, m_u, m_w)$, where

$$m_s = \frac{\sqrt{A_{sw}}\sqrt{A_{su}}}{\sqrt{A_{uw}}}, \quad m_u = \frac{\sqrt{A_{uw}}\sqrt{A_{su}}}{\sqrt{A_{sw}}}, \quad m_w = \frac{\sqrt{A_{sw}}\sqrt{A_{uw}}}{\sqrt{A_{su}}}. \tag{14}$$

Any negative component $A_{ij}$ in (10) under square roots has to be treated as:

$$\sqrt{A_{ij}} = i\sqrt{-A_{ij}}, \quad A_{ij} < 0, \tag{15}$$

where unit imaginary number is a factor on the right-hand side of (15). Because negative values always occur in pairs, the product or ratio of purely imaginary numbers is always real as also the $m_s$, $m_u$, $m_w$ components.

For $\gamma \cdot V \cdot \gamma$, from all the $V_{ij}$ components, it is convenient to find eigenvalues of $V$ by their invariants. Abbreviating

$$\begin{aligned} u_2 &= \text{Tr} V^2, \\ u_3 &= \text{Det} V, \end{aligned} \tag{16}$$

the eigenvalues as roots of the secular equation can be obtained using Cardano's method:

$$V_{kk} = \sqrt{\frac{2}{3}}\sqrt{u_2}\cos\left(\frac{\tau}{3} + \frac{2k\pi}{3}\right), \quad \cos\tau = 3\sqrt{6}\frac{u_3}{\sqrt{u_2^3}}. \tag{17}$$

In (16), $u_1=0$, since $V$ is a traceless tensor. For $k = 1,2,3$, $V_{kk}$ are the main components of the *efg* tensor in the *PAS*. The largest one in the absolute value is the $V_{zz}$. Remaining two are $V_{xx}$ and $V_{yy}$ and $\eta = (V_{xx}-V_{yy})/V_{zz}$, if we choose the axes so that $|V_{xx}| \leq |V_{yy}|$. We note that using (17), we determine eigenvalues or diagonal elements in *PAS* of the *efg*; however, orientation of the *PAS* with respect to the absorber frame is still undetermined.

Having the eigenvalues $V_{kk}$, one may construct projection operators as the Frobenius covariants $V^{(k)}$. Their explicit form in case of nondegenerate eigenvalues [10]:

$$V^{(k)} = \prod_{i \neq k} \frac{1}{V_{kk} - V_{ii}}(V - V_{ii}\mathbf{1}), \quad i, k = x, y, z, \tag{18}$$

allows construction of the eigenvectors. Because $V^{(k)}=e_k \otimes e_k$, where $e_k$ ($k=x,y,z$) is an orthonormal basis in which $V$ is diagonal, the components of $V^{(k)}$ operator determined in the absorber basis can be solved in the manner given in (14) to obtain explicit vectors forming *PAS* of the *efg*.

Equation (18) works for a nondegenerate case, that is, all three eigenvalues are different. For an axially symmetric *efg*, the component of $V$ along symmetry axis is $V_{axis}$ and the projection operator is $V^{(axix)}$:

$$\boldsymbol{V}^{(\text{axis})} = \frac{1}{3}\boldsymbol{1} + \frac{\text{sign}(u_3)}{\sqrt{3u_2}}\boldsymbol{V}, \quad V_{\text{axis}} = \frac{2}{\sqrt{3}}\text{sign}(u_3)\sqrt{u_2}. \quad (19)$$

Again, because $\boldsymbol{V}^{(\text{axix})} = \boldsymbol{e}_{\text{axis}} \otimes \boldsymbol{e}_{\text{axis}}$, using (15) one can determine unit vector $\pm\boldsymbol{e}_{\text{axis}}$ indicating direction of the symmetry axis. Equation (19) is valid only for a symmetric, traceless, second-order tensor in three dimensions.

Next, we discuss how to get texture-free spectra from a single crystalline absorber. First, let us consider the four directions of $\gamma$ parallel to the four directions pointing to vertices of a regular tetragon with its center at the origin. An example of these directions is (1,1,1), (–1,–1,1), (–1,1,–1), (1,–1,–1). One may check by direct calculations, that for any vector $\boldsymbol{m}$ and for any second-order, symmetric tensor $\boldsymbol{A}$, averaging of scalars $\boldsymbol{m}\cdot\boldsymbol{\gamma}$ or $\boldsymbol{\gamma}\cdot\boldsymbol{A}\cdot\boldsymbol{\gamma}$ over these four directions $\boldsymbol{\gamma}$, yields zero, which is independent of $\boldsymbol{\gamma}$. Therefore, this averaging results in spectra which do not depend on the direction of $\boldsymbol{\gamma}$, thus, we obtain texture-free spectra. Furthermore, in case of unpolarized radiation, the averaging can be accomplished using just three orthogonal directions. These considerations are consistent with group theory arguments and also with the concept of magic angle measurements [11, 12].

## 4. Half cube method of measurements

To determine all components of *efg* tensor and $\boldsymbol{B}$ vector, we need six measurements with $\boldsymbol{k}$ vector along the directions (1,0,0), (0,1,0), (0,0,1), (1,1,0), (1,0,1), (0,1,1). The half cube method, using a cubic sample holder shown in Fig. 1*b* is easy to perform. There are three holes with their axes being fourfold symmetry axes, and three other holes with their axes being twofold symmetry axes. The cube is divided into two halves by cutting along the plane shown by the black line in Fig. 1*b*. A planar absorber is mounted at the central position between the halves, with its plane parallel to the cross-section. The measurements are performed by orienting the holder such that during a particular measurement $\gamma$ rays are parallel to the axis of a hole.

A holder for texture-free measurements is shown in Fig. 1*c*. A cube with four holes along the directions (1,1,1), (1,1, –1), (1, –1,1), (–1,1, 1) is divided into two halves along a plane shown by the black line in Fig. 1*c*. An absorber is mounted between the two halves at the center of the cube.

## 5. Discussion and importance

The formalism shown here enables separation of *efg* and *hmf* in case of mixed interactions. The magnetic texture with respect to the absorber frame can be determined from the orientation of $\boldsymbol{B}$ axis. There is complete freedom to define the absorber reference frame.

Since equations (9–12) are linear functions of the velocity moments, for a single crystal with multiple sites, averaging of the tensors or vectors is implied if the subspectra cannot be distinguished.

The characteristic feature and advantage of the method discussed here is that one does not need perform any orientation of the single-crystal absorber. The absorber plate should be thin enough to be partially transparent for $\gamma$ radiation used in Mössbauer spectroscopy and to be cut parallel to the plane of arbitrary orientation. All derived formulas are valid for any orientation of a single crystal with respect to the absorber reference frame, and results (14, 18,

19) correspond to the components of vector ***B*** or tensor ***V*** in this arbitrary chosen reference frame *s*, *u*, *w* as shown in Fig. 1*a*.